\documentclass[prl,letterpaper,twocolumn,showpacs,superscriptaddress,amsmath,amsfonts,amssymb,floatfix,preprintnumbers,citeautoscript]
{revtex4}
\pdfoutput=1
\usepackage[T1]{fontenc}\usepackage[latin1]{inputenc}
\usepackage{dcolumn,graphicx,color,booktabs} \graphicspath{{Images/}}
\usepackage{amsmath,bm}
\usepackage[outercaption]{sidecap}
\renewcommand{\figurename}{Fig.}
\makeatletter\renewcommand{\fnum@figure}[1]{\figurename~\thefigure~(color online).}\makeatother
\definecolor{DarkBlue}{rgb}{0,0,0.5}

\begin{document} \pagestyle{plain}

\title{Fusion of bogoliubons in Ba$_{0.6}$K$_{0.4}$Fe$_2$As$_2$ and similarity of energy scales \\in high temperature superconductors}

\author{D.\,V.\,Evtushinsky}
\affiliation{Institute for Solid State Research, IFW Dresden, P.\,O.\,Box 270116, D-01171 Dresden, Germany}
\author{T.\,K.\,Kim}
\affiliation{Institute for Solid State Research, IFW Dresden, P.\,O.\,Box 270116, D-01171 Dresden, Germany}
\author{A.\,A.\,Kordyuk}
\affiliation{Institute for Solid State Research, IFW Dresden, P.\,O.\,Box 270116, D-01171 Dresden, Germany}
\affiliation{Institute of Metal Physics of National Academy of Sciences of Ukraine, 03142 Kyiv, Ukraine}
\author{V.\,B.\,Zabolotnyy}
\affiliation{Institute for Solid State Research, IFW Dresden, P.\,O.\,Box 270116, D-01171 Dresden, Germany}
\author{B.\,B\"{u}chner}
\affiliation{Institute for Solid State Research, IFW Dresden, P.\,O.\,Box 270116, D-01171 Dresden, Germany}
\author{A.\,V.~Boris}
\affiliation{Max-Planck-Institute for Solid State Research, Heisenbergstrasse 1, D-70569 Stuttgart, Germany}
\author{D.\,L.\,Sun}\author{C.\,T.~Lin}
\affiliation{Max-Planck-Institute for Solid State Research, Heisenbergstrasse 1, D-70569 Stuttgart, Germany}
\author{H.\,Q.\,Luo} \author{Z.\,S.\,Wang}\author{H.\,H.\,Wen}
\affiliation{Institute of Physics, Chinese Academy of Sciences, Beijing 100190, China}
\affiliation{National Laboratory of Solid State Microstructures and Department of Physics, Nanjing University, Nanjing 210093, China}
\author{R.\,Follath}
\affiliation{BESSY GmbH, Albert-Einstein-Strasse 15, 12489 Berlin, Germany}
\author{S.\,V.\,Borisenko}
\affiliation{Institute for Solid State Research, IFW Dresden, P.\,O.\,Box 270116, D-01171 Dresden, Germany}

\begin{abstract}

\noindent Usually the superconducting pairing is considered to modify electronic states only in a narrow momentum range close to the Fermi surface. Here we present a direct experimental observation of fusion of Bogoliubov dispersion branches originating from the antipodal Fermi crossings by means of angle-resolved photoemission spectroscopy (ARPES). Uncommon discernibility and brightness of bogoliubons' fusion stems from comparability of the superconducting gap magnitude and the distance from the Fermi level to the band's top, and strong electron scattering on a mode with similar energy. Such similarity of the electronic and pairing energy scales seems to be a persistent associate of high-temperature superconductivity (HTSC) rather than just a mere coincidence.

\end{abstract}


\maketitle

The most prominent signature of the superconductivity in the electronic spectrum is opening of the energy gap, $\Delta$, below critical temperature, $T_{\rm c}$, which is already inherent to the weak coupling BCS theory \cite{BCS}. Strong coupling between electrons and a mediator, indispensable for high $T_{\rm c}$, results in a more complex modification of the electronic spectrum at superconducting transition. Analysis of such modification allows for direct studies of electronic interactions with mediator spectrum \cite{Scalapino}. Noticeable spectroscopic evidence for the strong coupling is the depletion of the spectral weight below $T_{\rm c}$ at energies larger than $\Delta$, which was observed by means of tunneling and photoemission spectroscopy for Pb \cite{Giaever, Shin_Pb} and cuprate high temperature superconductors \cite{Kaminski, Sato}. Other hallmarks of the strong coupling superconductivity are large values of $2\Delta/k_{\rm B}T_{\rm c}$ and comparability of $\Delta$ and characteristic mediator energies, $\omega_{\rm m}$.

For iron-based superconductors, exhibiting $T_{\rm c}$ up to 56 K, the values of $2\Delta/k_{\rm B}T_{\rm c}$ reach 7 and higher \cite{Ding, PRB, NJP, Boris_cp}, unambiguously implying strong coupling regime. Also evidence for strong coupling of electrons to a bosonic spectrum were reported \cite{ding_mode, Boris_opt}. Here we present an observation of a peculiar evolution of the spectral function across the superconducting transition, detected by means of the angle-resolved photoemission spectroscopy (ARPES), carried on the highest-quality Ba$_{1-x}$K$_{x}$Fe$_2$As$_2$ (BKFA) single crystals with $T_{\rm c}$ of 38\,K. The most prominent and unusual characteristic of the observed behavior is the anomalously intense and well discernable fusion of the Bogoliubov dispersion branches \cite{Bogoliubov}, originating from the antipodal Fermi crossings\,---\,clear spectral weight emerges below $T_{\rm c}$ in the energy--momentum region where were nothing above $T_{\rm c}$ (Fig.~1). Such isolation of the large portion of the spectral weight, occurring with cooling through $T_{\rm c}$,  renders it suitable for detailed studies and suggests that this feature has to be captured in theoretical models for superconductivity in iron arsenides. Presented data not only emphasizes the similarity of $\Delta$ and characteristic energies of bosonic spectrum, but also shows the similarity of these two parameters to the electronic energy scale. The overview of the parameters of different materials suggests that in many high temperate superconductors $\Delta$ is comparable to the distance from the Fermi level (FL) to the nearest band structure peculiarity.

\begin{figure}[rb]\vspace{-0.2cm}
\includegraphics[width=0.6\columnwidth ]{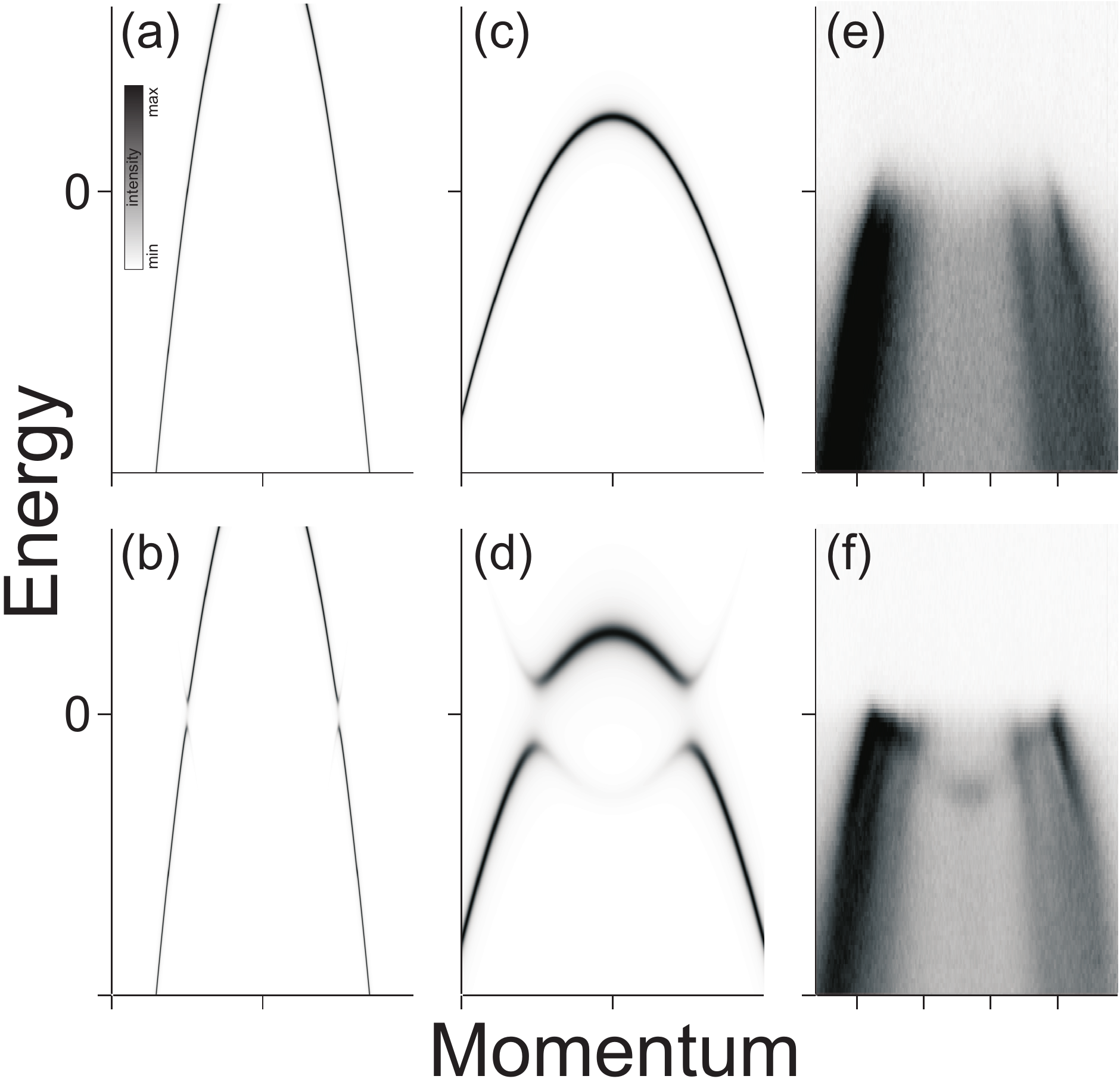}\vspace{-0.1cm}
\caption{(a), (b): Spectral function of an ordinary superconductor in the normal (a) and superconducting (b) state. The superconducting gap is much less than the characteristic depth of the electronic band. (c), (d): Spectral function in the case of comparable band depth and magnitude of the superconducting gap. (e), (f): ARPES spectra of optimally doped Ba$_{1-x}$K$_x$Fe$_2$As$_2$ ($T_{\rm c}=38$\,K) taken at $T=41$ and 1\,K respectively.}
 \label{f:Model1}
\end{figure}

\begin{figure*}[]
\includegraphics[width=\textwidth ]{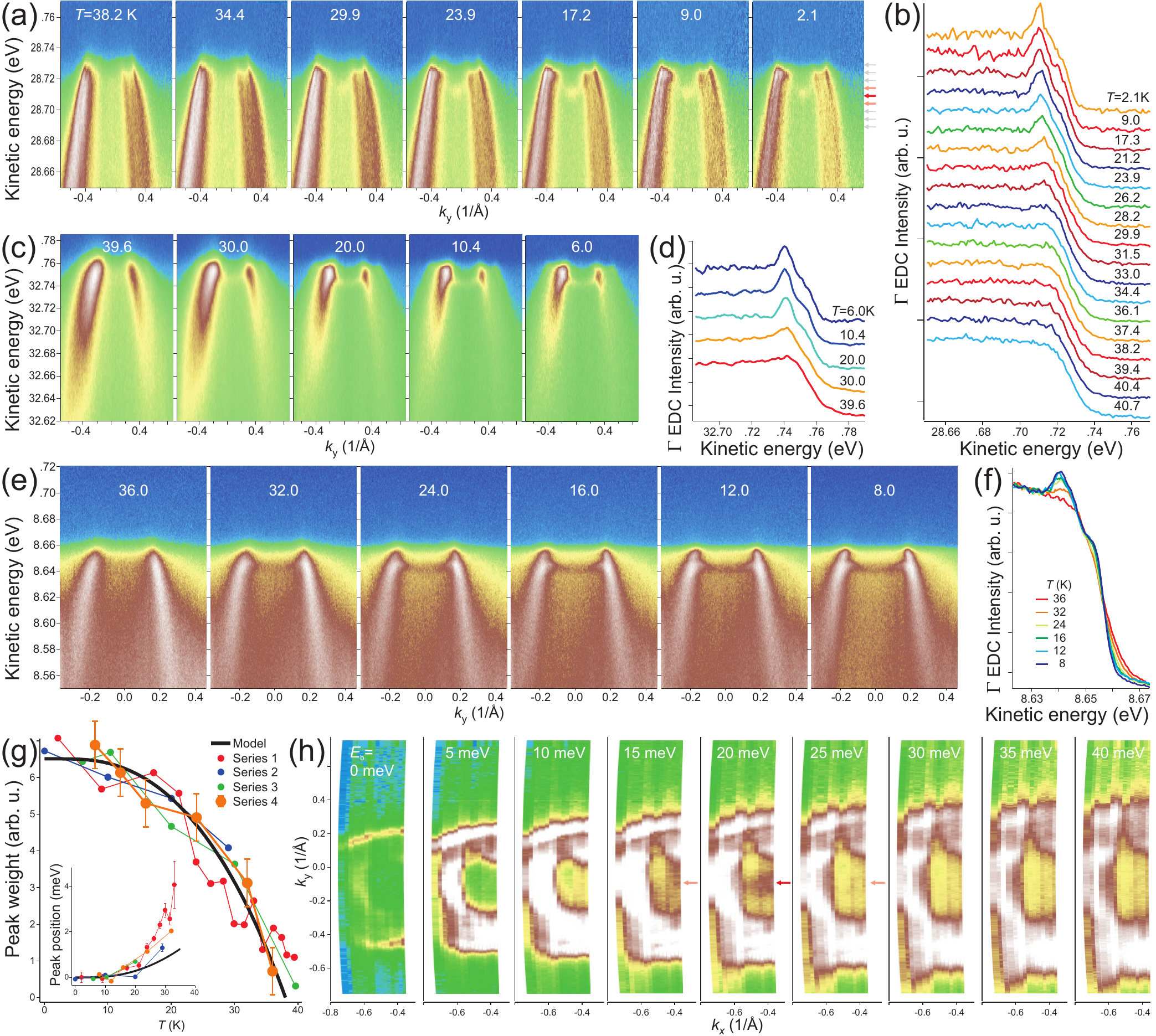}\vspace{-0.2cm}
\caption{Temperature dependence of the spectra near $\Gamma$ point. (a) Energy--momentum cut through the $\Gamma$ point, recorded at different temperatures. (b) Corresponding energy distribution curves (EDCs). (c,d) and (e,f) the same as (a,b) measured at different excitation energies. (g) Temperature dependence of the weight under the Bogoliubov peak: series\,1\,---\,cooling, data from (a,b); series\,2\,---\,$h\nu=35$\,eV, warming, data not shown; series\,3\,---\,cooling, data from (c,d); series\,4\,---\,warming, data from (e,f). Inset: temperature dependence of the peak position. (h) Constant energy cuts through the distribution of photoemission intensity recorded below 1\,K, integrated in the $\pm1$\,meV windows around the positions, indicated by arrows in the last panel of (a).}
 \label{f:Model1}
\end{figure*}

In Fig.~1(a),~(b) a modeled spectral function \cite{Spectr_func_SC} of an ordinary superconductor
is shown; typically in this case the superconducting gap is small, while the distance from the band's top to FL, $\varepsilon_0$, is large. Fig.~1(c),~(d) show the spectral function for the case of comparable $\varepsilon_0$ and $\Delta$\,---\,parameters that, as argued below, may be found in an unconventional superconductor. The energy--momentum cuts, recorded in the ARPES experiments on the optimally doped self-flux-grown Ba$_{1-x}$K$_x$Fe$_2$As$_2$, are shown in Fig.~1(e),~(f). The fusing branches of Bogoliubov dispersion are clearly seen below $T_{\rm c}$ as photoemission intensity between the normal state dispersion curves.

In order to address the origin of the found spectral feature, we have measured the temperature dependence of the cut passing through the center of the Brillouin zone (BZ) [Fig.~2(a)]. Energy distribution curves (EDC) directly from $\Gamma$ point are shown in Fig.~2(b), clearly revealing the development of the Bogoliubov peak when crossing $T_{\rm c}$. To ensure the robustness of the observation, in Fig.~2(c,d) and (e,f) equivalent data, taken from different samples with different excitation energies, are presented \cite{nonSC_surface}. Panel (h) shows the constant energy cuts through photoemission intensity distribution. Red arrow points to the rather symmetric intensity blob, centered at zero electron momentum, which corresponds to the bottom of the Bogoliubov dispersion, emphasizing that the fusion of bogoliubons is inherent to the entire two-dimensional spectral function. The temperature dependence of the weight under the peak in $\Gamma$-point EDC is shown in the Fig.~2(g); the same temperature dependence, calculated from the mentioned model \cite{Spectr_func_SC}, described by a simple expression $1/2-\frac{\varepsilon_0/2}{\sqrt{\varepsilon_0^2 + \Delta^2(T)}}$, well matches the experimental data. Within the same model we got an estimate $\varepsilon_0$=13\,meV. The inset to the Fig.\,2(g) shows the temperature dependence of the Bogoliubov peak position, which obviously departs from the model \cite{Spectr_func_SC}, showing that such a simple model is not qualitatively valid for all the cases.

\begin{figure}[]
\includegraphics[width=\columnwidth ]{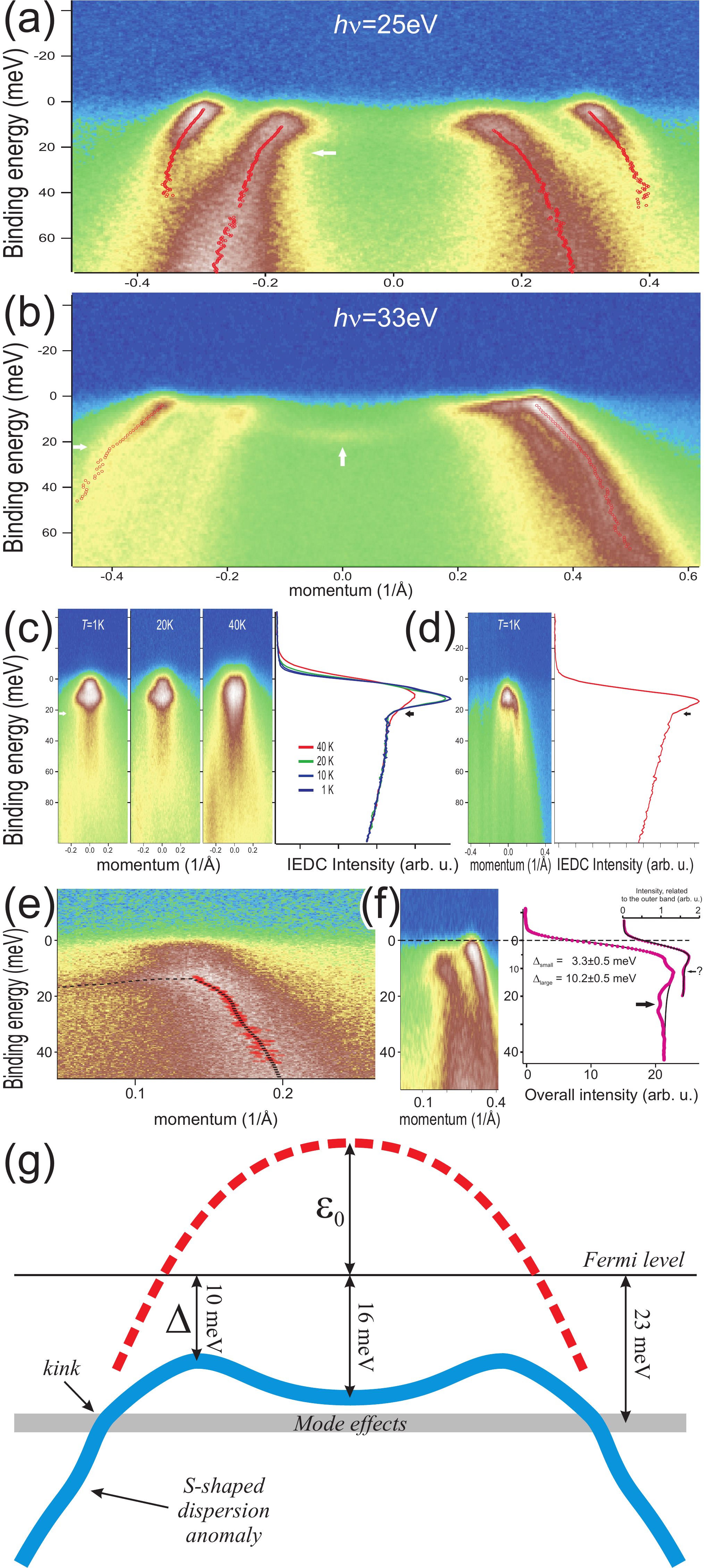}
\caption{(a, b) Energy--momentum cuts passing approximately through the $\Gamma$, recorded with $h\nu=25$\,eV (second BZ) at 7\,K,  and with $h\nu=33$\,eV at 1\,K respectively. (c) Temperature dependence of the cut and integrated energy distribution curve (IEDC) for electron-like X pocket (propeller's shaft). (d) Propeller's blade at 1\,K. (e) Zoomed in dispersion of the inner $\Gamma$ barrel at 7\,K, revealing \textit{S}-shaped dispersion in addition to the kink. (f) Determination of the superconducting gap for $\Gamma$ bands via fitting of IEDC. (g) Empirical model for dispersion of the inner $\Gamma$ band in the superconducting (blue) and normal (red) state.}
 \label{f:Model1}
\end{figure}

Now we take a closer look at the structure of the spectral function on the relevant energy scale in the whole Brillouin zone. The FS of BKFA consists of two $\Gamma$-barrels, and a propeller-like structure at the corner of BZ \cite{Volodya, NJP}. Earlier it was found that the superconducting gap is large (2$\Delta_{\rm large}/kT_{\rm c}\simeq6.5$) for all FS sheets except for the outer $\Gamma$ barrel \cite{PRB, NJP, Ding}. Presented here experimental data confirms those conclusions and allows for improvement of the estimate for the smaller gap, $\Delta_{\rm small}=3.3\pm0.5$\,meV (2$\Delta_{\rm small}/kT_{\rm c}\simeq2.0$) [Fig.~3(f)], which is in detailed agreement with specific heat measurements \cite{Boris_cp}. ARPES data, revealing effects of the mode scattering on the electronic spectrum of BKFA are shown in Fig.~3: kinks at around 23\,meV binding energy in the band dispersion for both inner \cite{3rd_band} and outer $\Gamma$ bands are well discernable in the data, taken deeply in the superconducting state [Fig.~3(a,b,e)]; depletion of the spectral weight at the same energy develops upon cooling through $T_{\rm c}$ [Fig.~3(c,d)]; the \textit{S}-shaped dispersion is observed on the inner $\Gamma$ band [Fig.~3(e)]. Thus, strong coupling to a mode is detected at low temperatures for all bands at the FL around binding energy of 23\,meV, which results from the coupling of the electronic spectrum, gapped with $\Delta_{\rm large}=10$\,meV, to a mode with energy $\Omega_{\rm M}\simeq13$\,meV, very close to the energy of the resonance peak ($\sim14$\,meV), observed in neutron scattering \cite{Christianson, Boris_opt}. In addition, some peculiarities are present on the outer $\Gamma$ band at binding energies below 20\,meV [Fig.~3(f)], maybe originating from scattering of the part of the spectrum gapped with $\Delta_{\rm small}$.

The most interesting mode-related observation here is that the bottom of the bogoliubons' fusion is situated just above the energy, where effects of the mode are located, suggesting that \emph{Bogoliubov dispersion is squeezed and shifted towards the FL due to interaction with the mode}. At the same time it seems that the bogoliubons' fusion owes its high intensity not only to the interaction with the mode, but also to the close location of the band's top. Though the estimate for the distance from the top of the inner $\Gamma$-barrel to the FL, obtained above, can be less than the actual value, an independent estimate for $\varepsilon_0$, derived from the Fermi-function-divided high temperature ARPES spectra, yields 13 to 25\,meV for $\varepsilon_0$ (uncertainty comes from temperature broadening of the spectra) is still close to $\Delta$ and $\Omega_{\rm M}$.

\begin{figure}[]
\includegraphics[width=\columnwidth ]{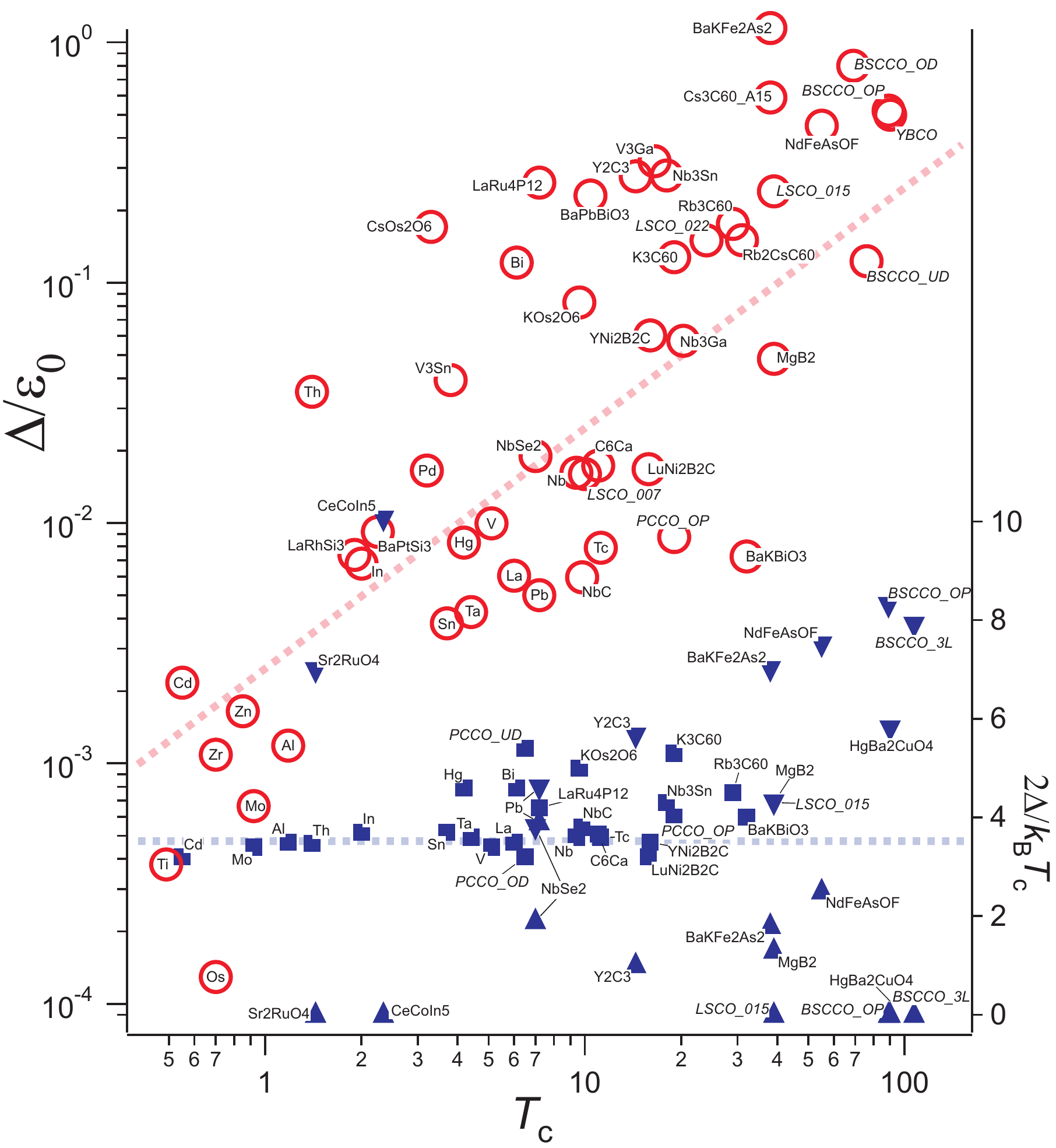}\vspace{-0.3cm}
\caption{Ratio of the pairing energy, $\Delta$, to the characteristic electronic energy, namely, distance from the Fermi level to the nearest flat region of the band dispersion, $\varepsilon_0$,  as a function of $T_{\rm c}$ (red circles, left axis). BCS ratio, $2\Delta/k_{\rm B}T_{\rm c}$, as a function of $T_{\rm c}$ (blue squares and triangles, right axis). Pink dashed line corresponds to $\Delta/\varepsilon_0\sim T_{\rm c}$, horizontal light blue dashed line indicates $2\Delta/k_{\rm B}T_{\rm c} = 3.53$. Corresponding references can be found in the supplementary materials.}
 \label{f:Model1}
\end{figure}

Moreover, the distance to the FL from the top (bottom) of the propeller bands is even smaller, 5--15\,meV \cite{PRB, NJP, Volodya}, making clear that all three energies, $\Delta$, $\Omega_{\rm M}$, and $\varepsilon_0$ are very close in BKFA. Is such a situation unique? In Fig.~4 the ratio $\Delta/\varepsilon_0$ is plotted versus $T_{\rm c}$ for most studied compounds. One might see that at present many of superconductors with highest $T_{\rm c}$ have $\Delta$ comparable to $\varepsilon_0$, therefore when analyzing experimental data one should be aware of this fact\,---\,for instance, van Hove singularity, situated close to the Fermi level, can be mistaken for the (pseudo)gap in spectroscopic methods, and result in unusual temperature dependence of transport, thermal, and other physical properties. Theoretically the most simple case of $\Delta\ll\omega_{\rm m}\ll\varepsilon_0$ is considered in the original BCS approach. Later it was shown, both theoretically and experimentally, that in the case of strong coupling, when $\Delta$ becomes comparable to $\omega_{\rm m}$, the BCS ratio $2\Delta/k_{\rm B}T_{\rm c}$ substantially grows above the value of 3.53 \cite{Carbotte} (Fig.~4). The case of $\omega_{\rm m}\sim\varepsilon_0$ so far attracted much less attention \cite{Grimaldi}, while presented here data analysis suggests that for high-$T_{\rm c}$ superconductors $\Delta\sim\omega_{\rm m}\sim\varepsilon_0$.
\vspace{-0.1cm}



In conclusion, we presented a direct observation of entire Bogoliubov dispersion branch, usually inaccessible experimentally. The found spectral feature offers a possibility of detailed studies of electronic interactions in iron-based superconductors. Abnormal brightness of bogoliubons' fusion is powered by comparability of all relevant energy scales\,---\,electronic band energy, pairing energy, and energy of a mode.\\
\indent
We acknowledge helpful discussions with D.\,S.\,Inosov, A.\,V.\,Chubukov, and A.\,A.\,Yaresko and technical support from R.\,H\"{u}bel.



\begin{thebibliography}{00}
\bibitem{nonSC_surface} Some data [e.g. Fig.~2(e), Fig.3~(c,e)] exhibit presence of non-superconducting component. Most likely in this case it comes from the aged surface layer.
\end{thebibliography}
\end{document}